\documentclass[aps,prl,reprint,groupedaddress]{revtex4-1}
\usepackage{graphicx} 
\usepackage{dcolumn} 
\usepackage{bm}
\usepackage{hyperref}

\begin{document}

\title{Isospin dynamics on neck fragmentation in isotopic nuclear reactions}
\author{Zhao-Qing Feng$^{1,2}$}
\email{fengzhq@impcas.ac.cn}

\affiliation{$^{1}$Institute of Modern Physics, Chinese Academy of Sciences, Lanzhou 730000, People's Republic of China            \\
$^{2}$Kavli Institute for Theoretical Physics, Chinese Academy of Sciences, Beijing 100190, People's Republic of China}

\date{\today}

\begin{abstract}
The neck dynamics in Fermi-energy heavy-ion collisions, to probe the nuclear symmetry energy in the domain of sub-saturation densities, is investigated within an isospin dependent transport model. The single and double ratios of neutron/proton from free nucleons and light clusters (complex particles) in the isotopic reactions are analyzed systematically. Isospin effects of particles produced from the neck fragmentations are explored. It is found that the ratios of the energetic isospin particles strongly depend on the stiffness of nuclear symmetry energy and the effects increase with softening the symmetry energy, which would be a nice probe for extracting the symmetry energy below the normal density in experimentally. A flat structure appears at the tail spectra from the double ratio distributions. The neutron to proton ratio of light intermediate mass fragments (IMFs) with charged number Z$\leq$8 is related to the density dependence of symmetry energy with less sensitivity in comparison to the isospin ratios of nucleons and light particles.

\begin{description}
\item[PACS number(s)]
21.65.Ef, 24.10.Lx, 25.75.-q
\end{description}
\end{abstract}

\maketitle

Heavy-ion collisions at the Fermi energies (10-100\emph{A} MeV) attract much attention on several topical issues in nuclear physics, i.e., spinodal multifragmentation, liquid-gas phase transition, properties of highly excited nuclei, symmetry energy in the domain of subnormal densities etc \cite{Ch04,Co93,Po95,Wu98,Ma99}. The isospin dynamics in the Fermi-energy heavy-ion collisions is related to several interesting topics, e.g., in-medium nucleon-nucleon cross sections, cluster formation, isotopic distribution, the symmetry energy in dilute nuclear matter etc. It has been obtained much progress in constraining the density dependence of symmetry energy in the domain of sub-saturation densities both in theories \cite{Li97,Ch05,Ba05,Li08,Ts09,Zh14} and in experiments \cite{Li04,Fa06,Xi06,Su10}. Different mechanisms can coexist and are correlated in heavy-ion collisions at the Fermi energies, such as projectile or target fragmentation, neck emission, preequilibrium emission of light clusters (complex particles), fission of heavy fragments, multifragmentation etc, in which the isospin dependent nucleon-nucleon potential dominates the dynamical processes. The time scales from dynamical and pre-equilibrium emissions to statistical decay of excited systems at equilibrium and the isospin effect in neck fragmentation were investigated in experimentally \cite{Fi14}. Transport models are needed to understand the mechanisms correlated with the emission time and the dependence on incident energy, collision centrality and reaction systems.

In this work, the Lanzhou quantum molecular dynamics (LQMD) model \cite{Fe11} is used to investigate the isospin dynamics in the Fermi-energy domain for the first time. The model has been used to extract the density dependence of the nuclear symmetry energy from heavy-ion collisions, in particular in the domain of high-baryon densities. Several promising observables have been proposed for measurements in experiments \cite{Fe12,Fe13}. In the model, all possible reaction channels in charge-exchange processes, elastic and inelastic collisions by distinguishing isospin effects were included \cite{Fe15}. The temporal evolutions of the baryons (nucleons and resonances) and mesons in the reaction system under the self-consistently generated mean-field are governed by Hamilton's equations of motion. Based on the Skyrme interactions, we constructed an isospin, density and momentum-dependent potentials originated from the Hamiltonian, which consists of the relativistic energy, the effective interaction and the momentum-dependent potentials. The effective interaction potential is composed of the Coulomb potential and the local interaction.

The local interaction potential is derived from the energy-density functional as the form of
$U_{loc}=\int V_{loc}(\rho(\mathbf{r}))d\mathbf{r}$. The functional reads
\begin{eqnarray}
V_{loc}(\rho)=&& \frac{\alpha}{2}\frac{\rho^{2}}{\rho_{0}} +
\frac{\beta}{1+\gamma}\frac{\rho^{1+\gamma}}{\rho_{0}^{\gamma}} + E_{sym}^{loc}(\rho)\rho\delta^{2}
\nonumber \\
&& + \frac{g_{sur}}{2\rho_{0}}(\nabla\rho)^{2} + \frac{g_{sur}^{iso}}{2\rho_{0}}[\nabla(\rho_{n}-\rho_{p})]^{2},
\end{eqnarray}
where the $\rho_{n}$, $\rho_{p}$ and $\rho=\rho_{n}+\rho_{p}$ are the neutron, proton and total densities, respectively, and the $\delta=(\rho_{n}-\rho_{p})/(\rho_{n}+\rho_{p})$ being the isospin asymmetry. The parameters $\alpha$, $\beta$ and $\gamma$ are taken to be -226.5 MeV, 173.7 MeV and 1.309, respectively. The set  of the parameters gives the compression modulus of K=230 MeV for isospin symmetric nuclear matter at the saturation density ($\rho_{0}=0.16$  fm$^{-3}$). The surface coefficients $g_{sur}$ and $g_{sur}^{iso}$ are taken to be 23 MeV fm$^{2}$ and -2.7 MeV fm$^{2}$, respectively. It should be noticed that the momentum-dependent interaction \cite{Fe11} is not included in this work because we concentrate on the isospin dynamics from the neck fragmentations in the Fermi-energy heavy-ion collisions, in which the reaction systems need to evolve over 500 fm/c, e.g., 35 MeV/nucleon in this work. The momentum-dependent potential shortens the stability in the initialization for constructing a ground-state nucleus.

\begin{figure}
\includegraphics[width=8 cm]{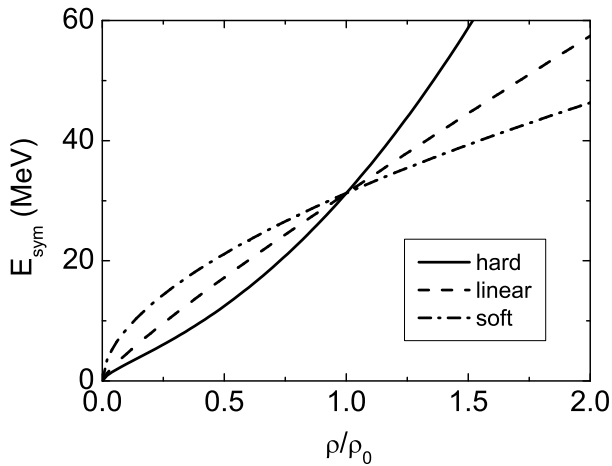}
\caption{\label{fig:epsart} Nuclear symmetry energy as a function of baryon density for the soft, linear and hard cases denoted by solid, dashed and dot-dashed lines, respectively.}
\end{figure}

The symmetry energy is composed of the kinetic energy from the Fermi motions of nucleons and the local density-dependent mean-field potentials as
\begin{equation}
E_{sym}(\rho)=\frac{1}{3}\frac{\hbar^{2}}{2m}\left(\frac{3}{2}\pi^{2}\rho\right)^{2/3}+E_{sym}^{loc}(\rho).
\end{equation}
The local part is adjusted to mimic predictions of the symmetry energy calculated by microscopical or phenomenological many-body theories and has the form of
\begin{equation}
E_{sym}^{loc}(\rho)=\frac{1}{2}C_{sym}(\rho/\rho_{0})^{\gamma_{s}}.
\end{equation}
The parameter $C_{sym}$ is taken as the value of 38 MeV. We name the $\gamma_{s}$=0.5, 1 and 2 being the soft, linear and hard symmetry energy, respectively, which cross at saturation density with the value of 31.5 MeV. Shown in Fig. 1 is a comparison of different stiffness of symmetry energy. It should be noticed that the symmetry energy contributes an opposite isospin dynamics in heavy-ion collisions in the low-density region in comparison to the ones at supra-saturation densities.

\begin{figure*}
\includegraphics[width=16 cm]{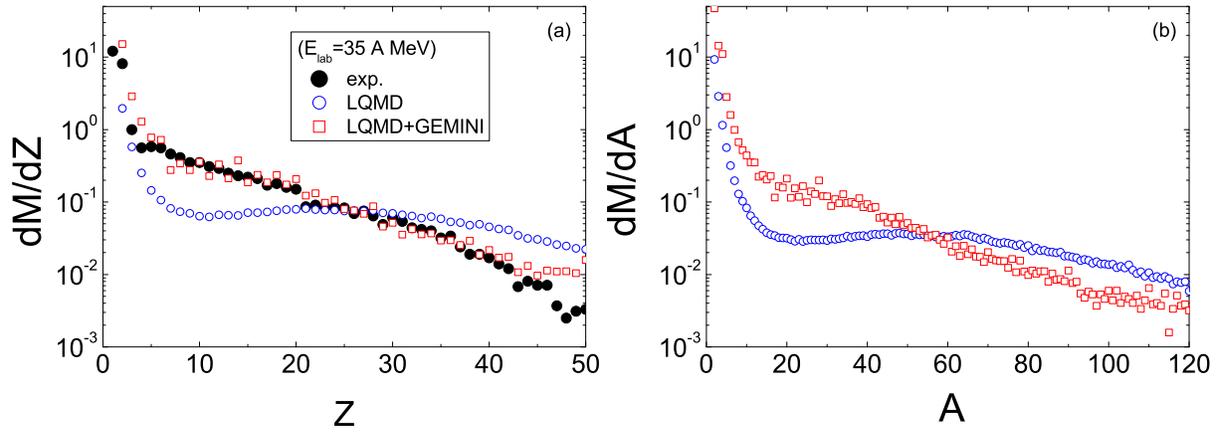}
\caption{\label{fig:wide}(Color online) Fragment distributions as functions of charged and mass numbers in central $^{197}$Au+$^{197}$Au collisions.}
\end{figure*}

The composite system formed in Fermi energy heavy-ion collisions is highly excited with an excitation energy up to several tens of MeV/nucleon. The heated system is unstable and fragmentation even multifragmentation takes place. As a test of the model, the fragmentation reactions in central $^{197}$Au+$^{197}$Au collisions at an incident energy of 35 MeV/nucleon have been investigated as shown in Fig. 2. The available data from MSU-NSCL \cite{De98} can be reproduced nicely well from the LQMD transport model combined with the GEMINI statistical decay code for excited fragments \cite{Ch88}. The fragmentation dynamics in the Fermi-energy heavy-ion collisions is described by the LQMD model. The primary fragments are constructed in phase space with a coalescence model, in which nucleons at freeze-out are considered to belong to one cluster with the relative momentum smaller than $P_{0}$ and with the relative distance smaller than $R_{0}$ (here $P_{0}$ = 200 MeV/c and $R_{0}$ = 3 fm). At the freeze-out, the primary fragments are highly excited. The de-excitation of the fragments is described within the GEMINI code. It is noticed that a flat structure appears from the spectra of primary fragments. The yields of IMFs (3$\leq Z \leq$30) are underestimated. However, heavier fragments ($Z > $30, $A > $60) are overestimated in the charge and mass distributions. It is caused from the fact that the ability of fluctuation is limited for the cold fragment formation. The combined approach enhances the production of IMFs, but reduces the heavier fragments by the de-excitation process with larger excitation energies.

\begin{figure*}
\includegraphics[width=16 cm]{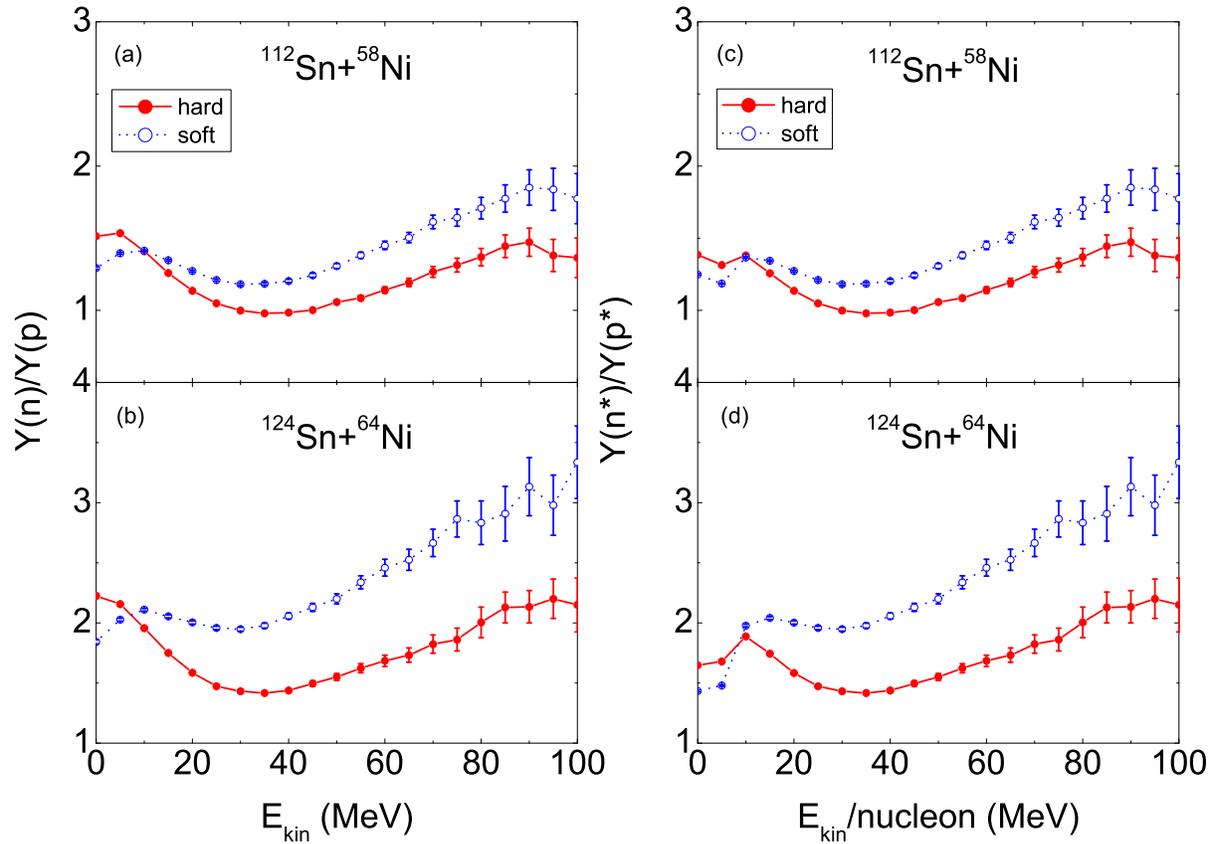}
\caption{\label{fig:wide}(Color online) Kinetic energy spectra of neutron/proton ratios from the yields of free nucleons and 'gas-phase' nucleons (nucleons, hydrogen and helium isotopes) from neck fragmentations in the $^{112}$Sn+$^{58}$Ni and $^{124}$Sn+$^{64}$Ni reactions at the fermi energy of 35 MeV/nucleon within the collision centralities of 6-8 fm.}
\end{figure*}

\begin{figure*}
\includegraphics[width=16 cm]{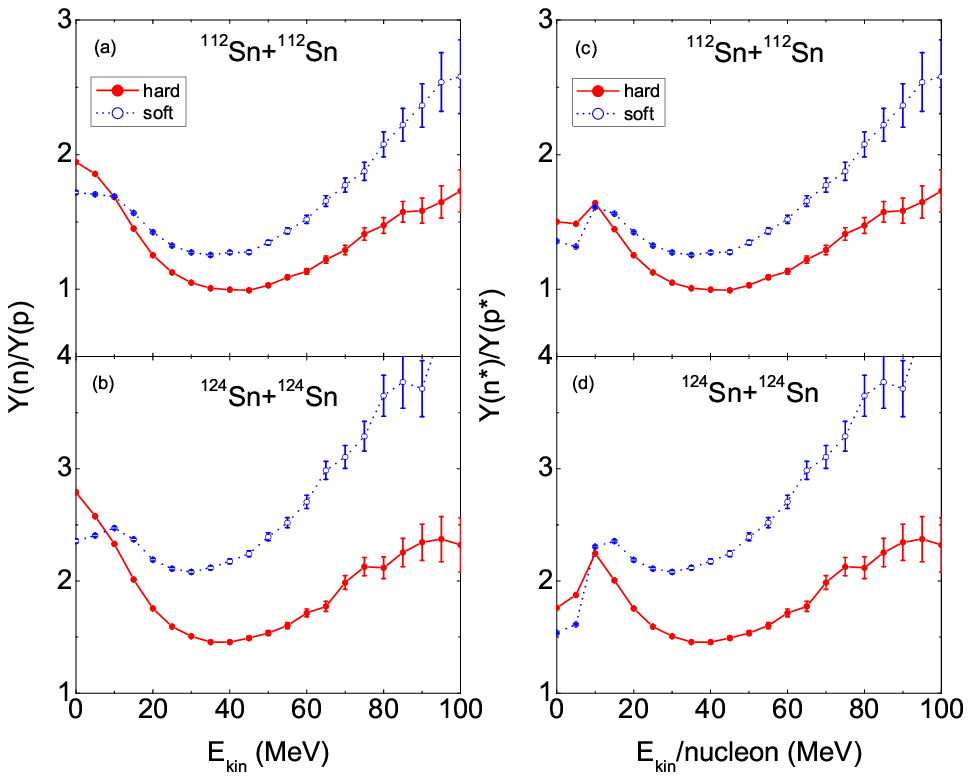}
\caption{\label{fig:wide}(Color online) The same as in Fig. 3, but for the reactions of $^{112}$Sn+$^{112}$Sn and $^{124}$Sn+$^{124}$Sn.}
\end{figure*}

The dynamics of preequilibrium nucleons and light fragments produced in heavy-ion collisions is influenced by the mean-field potentials. To extract the sub-saturation density information of nuclear symmetry energy, one needs to produce the observables formed in the low-density domain. Particles produced from the neck fragmentations in heavy-ion collisions could be probes of the low-density phase diagram, which are constrained within the midrapidities ($|y/y_{proj}|<$0.3) in semicentral nuclear collisions. Shown in Fig. 3 is the kinetic energy spectra of neutron/proton ratios from the yields of free nucleons and 'gas-phase' nucleons (nucleons, hydrogen and helium isotopes) from the neck fragmentations in the $^{112}$Sn+$^{58}$Ni and $^{124}$Sn+$^{64}$Ni reactions at a beam energy of 35 MeV/nucleon with the different symmetry energies. One can see that the ratios decrease with the kinetic energy. After a turning point around the Fermi energy (36 MeV), the value is enhanced. The structure of the spectra is determined by the competition of Coulomb interaction between charged particles and symmetry potential, which enhance the proton and neutron yields, respectively. The symmetry energy effect is pronounced and close to 50$\%$ because of the longer isospin relaxation time in Fermi-energy heavy-ion collisions in comparison to the high-density observables, such as the n/p ratio at midrapidities, $\pi^{-}/\pi^{+}$, $K^{0}/K^{+}$, $\Sigma^{-}/\Sigma^{+}$ etc \cite{Fe12,Fe13}. A larger value of the n/p ratio with softening symmetry energy is found, which is caused from the fact that more repulsive interaction enforced on neutrons. Similar conclusions are drawn in the isotopic reactions of $^{112}$Sn+$^{112}$Sn and $^{124}$Sn+$^{124}$Sn as shown in Fig. 4. More pronounced isospin effect is found in the neutron-rich systems. The results are consistent with the ratio of pre-equilibrium neutrons to protons within the isospin-dependent Boltzmann-Uehling-Uhlenbeck (IBUU) transport model calculations \cite{Li97}.

\begin{figure*}
\includegraphics[width=16 cm]{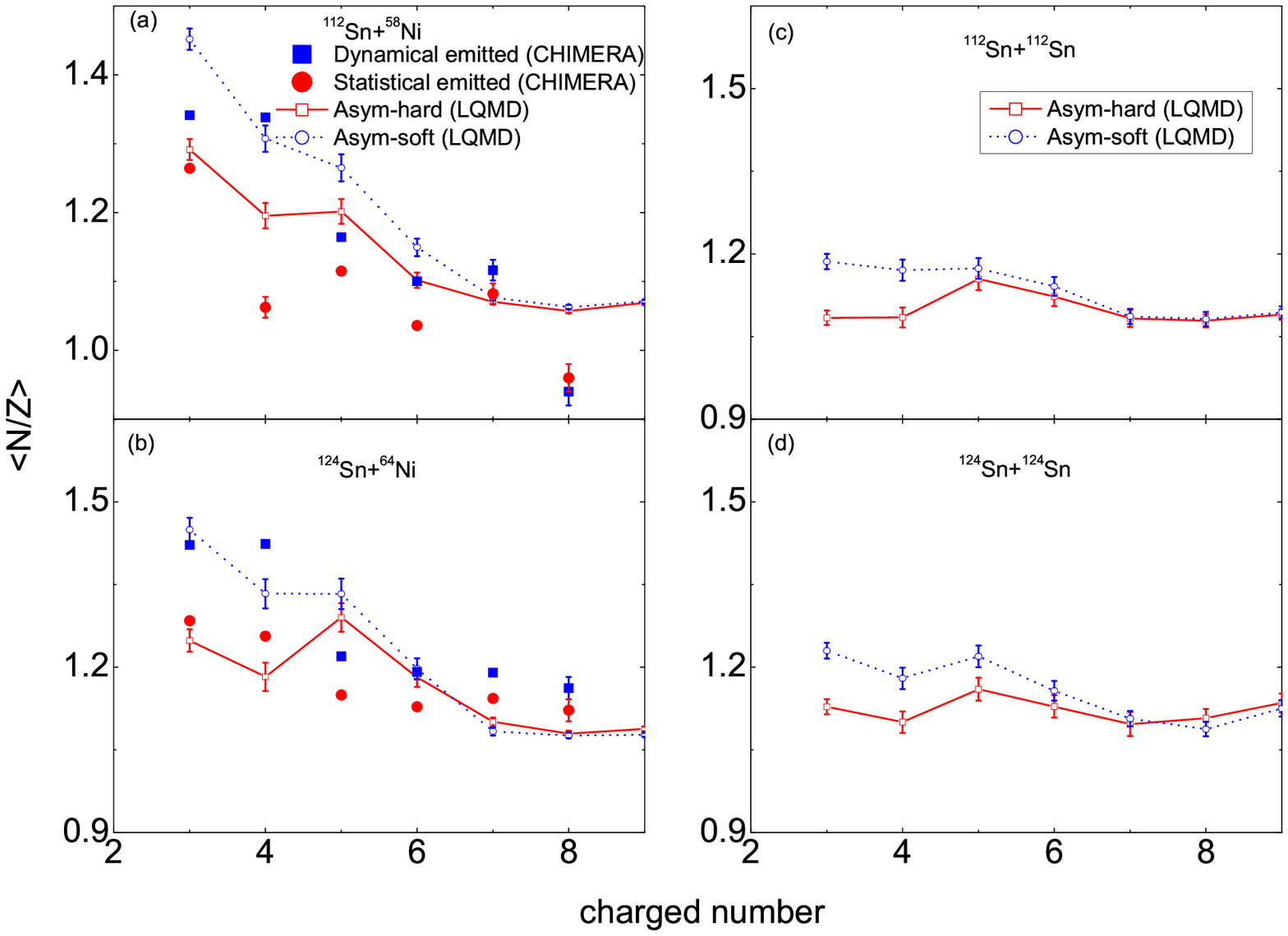}
\caption{\label{fig:wide}(Color online) The average neutron to proton ratio of IMFs emitted within the rapidity range of $|y/y_{proj}|<$0.3 as a function of charged number with the different symmetry energies. The data from CHIMERA detector \cite{Fi12} are shown for comparison.}
\end{figure*}

Besides the fast nucleons being probes of symmetry energy in the dilute matter, the neutron to proton ratio of light intermediate mass fragments (IMFs) could be sensitive to the stiffness of symmetry energy because of the isospin migration from the neck fragmentation \cite{Ba04}. The light IMFs (Z$\leq$10) are measured by the CHIMERA detector at the INFN-LNS Superconducting Cyclotron of Catania (Italy), and emitted preferentially towards the midrapidity domain on a short timescale within 50 fm/c, which can not be entirely described through the decay of the excited projectile-like (PLF) and target-like (TLF) fragments. We constrained the particles emitted from the neck fragmentation within the rapidity range of $|y/y_{proj}|<$0.3. Shown in Fig. 5 is the average neutron to proton ratio of light IMFs in the isotopic reactions of $^{112}$Sn+$^{58}$Ni, $^{124}$Sn+$^{64}$Ni, $^{112}$Sn+$^{112}$Sn and $^{124}$Sn+$^{124}$Sn. The isospin effect is pronounced for the light isotopes, i.e., lithium, beryllium and boron. The soft symmetry energy leads to larger n/p ratio of light IMFs.

\begin{figure*}
\includegraphics[width=16 cm]{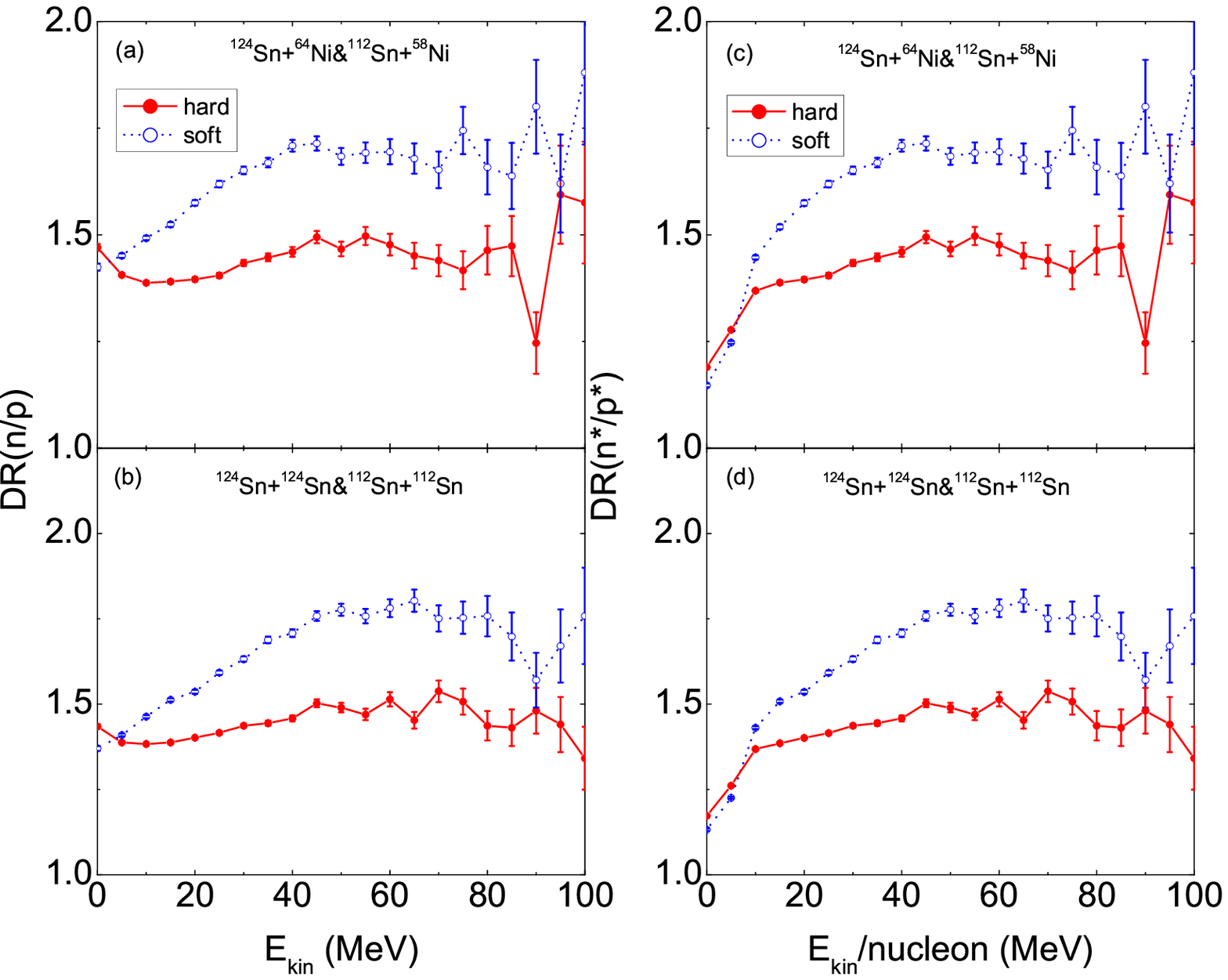}
\caption{\label{fig:wide}(Color online) The double neutron/proton ratios of free nucleons and 'gas-phase' particles in collisions of $^{124}$Sn+$^{64}$Ni over $^{112}$Sn+$^{58}$Ni and $^{124}$Sn+$^{124}$Sn over $^{112}$Sn+$^{112}$Sn at an incident energy of 35 MeV/nucleon and with the impact parameters of 6-8 fm. }
\end{figure*}

In order to reduce the influence of the Coulomb force and less systematic errors, the kinetic energy spectra of double ratios have been analyzed, which are defined as the ratios of isospin observables (n/p, $\pi^{-}/\pi^{+}$, $K^{0}/K^{+}$ etc) from two isotopic systems. We calculated the double neutron/proton ratios of free nucleons and 'gas-phase' particles in the semicentral collisions of $^{124}$Sn+$^{64}$Ni over $^{112}$Sn+$^{58}$Ni and $^{124}$Sn+$^{124}$Sn over $^{112}$Sn+$^{112}$Sn at an incident energy of 35 MeV/nucleon as shown in Fig. 6. One notices that the symmetry energy results in a variation of the ratio about 16$\%\sim$18$\%$, which is roughly twice in comparison to the high-energy heavy-ion collisions \cite{Fe12}. A smooth structure appears in the domain of high kinetic energies and very similar spectra for the four cases. The double ratios of preequilibrium nucleons and light clusters in the Sn isotopic reactions at the energies of 50 MeV/nucleon and 120 MeV/nucleon were measured at the National Superconducting Cyclotron Laboratory (NSCL), respectively \cite{Co14}. The calculations from the improved quantum molecular dynamics model and the isospin-dependent Boltzmann-Uehling-Uhlenbeck show that the double ratio spectra are related to the momentum-dependent interactions, in particular to the isovector part, but weakly depend on the stiffness of symmetry energy \cite{Zh14,Ko15}. At lower incident energies, the contribution of the momentum-dependent potential to the nucleon effective mass is negligible. The effects of symmetry energy on the double ratio spectra can not be changed by reaction systems and different observables at the considered energy.

In summary, within the LQMD transport model we have investigated the neck dynamics in Fermi-energy heavy-ion collisions, in which the single and double ratios of neutron to proton yields from free nucleons and light clusters as well as the neutron to proton ratio of light IMFs are particularly concentrated. The secondary decay of primary fragments formed in heavy-ion collisions increases the IMF production. The light IMFs from the neck fragmentation are emitted preferentially towards the midrapidity domain on a short timescale in comparison to PLFs and TLFs. The isospin ratios depend on the stiffness of symmetry energy and the effects increase with softening the symmetry energy, in particular in neutron-rich nuclear reactions.

I would like to thank Lie-Wen Chen, Maria Colonna and Jun Xu for fruitful discussions. This work was supported by the Major State Basic Research Development Program in China (2015CB856903), the National Natural Science Foundation of China Projects (Nos 11175218 and U1332207) and the Youth Innovation Promotion Association of Chinese Academy of Sciences.

\end{document}